\documentclass[12pt,letterpaper]{article} 
\usepackage{amssymb,amsmath}
\usepackage{graphicx} 
\usepackage{enumerate}
\usepackage{natbib}
\newcommand{\blind}{0}

\newcommand{\ave}{\mathop{\mbox{ave}}}
\newcommand{\med}{\mathop{\mbox{med}}}

\newcommand{\bzero}{\boldsymbol 0}

\newcommand{\bv}{\boldsymbol v}
\newcommand{\bx}{\boldsymbol x}

\newcommand{\bmu}{\boldsymbol \mu}

\newcommand{\bS}{\boldsymbol S}
\newcommand{\bT}{\boldsymbol T}
\newcommand{\bU}{\boldsymbol U}
\newcommand{\bX}{\boldsymbol X}
\newcommand{\bZ}{\boldsymbol Z}
\newcommand{\bSigma}{\boldsymbol \Sigma}

\newcommand{\bhmu}{\hat{\boldsymbol \mu}}
\newcommand{\bhSigma}{\hat{\bSigma}}

\def\spacingset#1{\renewcommand{\baselinestretch}
{#1}\small\normalsize} \spacingset{1}

\begin{document}
\if0\blind
{ \vspace{-1cm}
  \title{\bf Detecting Deviating Data Cells}
  \author{Peter J. Rousseeuw\thanks{
    The research of P. Rousseeuw has been supported by 
    projects of Internal Funds KU Leuven.
    W. Van den Bossche obtained financial support from the EU 
    Horizon 2020 project SCISSOR: Security in trusted SCADA 
    and smart-grids 2015--1017.
    The authors are grateful for interesting discussions with
    Mia Hubert and the participants of the BIRS Workshop 15w5003
		in Banff, Canada, November 16-20, 2015, and to
		Jakob Raymaekers for assistance with the CRAN package
		{\it cellWise}.
		The reviewers provided helpful suggestions to improve
		the presentation.} \hspace{.1cm}
    and 
    Wannes Van den Bossche\\
    Department of Mathematics, KU Leuven, Belgium}
	\date{June 4, 2017}
  \maketitle
} \fi

\if1\blind
{
  \bigskip
  \bigskip
  \bigskip
  \begin{center}
    {\LARGE\bf Detecting Deviating Data Cells}
  \end{center}
  \medskip
	\begin{center}
		{June 4, 2017}
	\end{center}
	\medskip
} \fi

\bigskip
\begin{abstract}
A multivariate dataset consists of $n$ cases in $d$ 
dimensions, and is often stored in an $n$ by $d$ data matrix.
It is well-known that real data may contain outliers. 
Depending on the situation, outliers may be (a) undesirable 
errors which can adversely affect the data analysis, or (b) 
valuable nuggets of unexpected information. 
In statistics and data analysis the word outlier usually 
refers to a row of the data matrix, and the methods to detect
such outliers only work when at least half the rows are clean.
But often many rows have a few contaminated cell values, 
which may not be visible by looking at each variable (column)
separately.
We propose the first method to detect deviating data cells 
in a multivariate sample which takes the correlations between 
the variables into account.
It has no restriction on the number of clean rows,
and can deal with high dimensions.
Other advantages are that it provides predicted values of the outlying cells, while imputing 
missing values at the same time.
We illustrate the method on several real data sets, where
it uncovers more structure than found by purely columnwise 
methods or purely rowwise methods.
The proposed method can help to diagnose why a certain
row is outlying, e.g. in process control.
It also serves as an initial step for estimating 
multivariate location and scatter matrices.
\end{abstract}

\vspace{0.3cm}

\vfill

\newpage
\spacingset{1.45} 

\section{\normalsize{INTRODUCTION}}
\label{sec:introduction}
Most data sets come in the form of
a rectangular matrix $\bX$
with $n$ rows and $d$ columns, where $n$ is called
the sample size and $d$ the dimension. 
The rows of $\bX$ correspond to the cases, whereas the 
columns are the variables.
Both $n$ and $d$ can be large, and it may happen that 
$d > n$. 
There exist many data models as well as techniques to fit 
them, such as regression and principal component analysis.

It is well-known that many data sets contain outliers. 
Depending on the circumstances, outliers may be (a) undesirable 
errors which can adversely affect the data analysis, or (b) 
valuable nuggets of unexpected information. 
Either way, it is important to be able to {\it detect} the 
outliers, which can be hard for high $d$.

\begin{figure}[!ht]
\centering
\includegraphics[width=10.3cm]{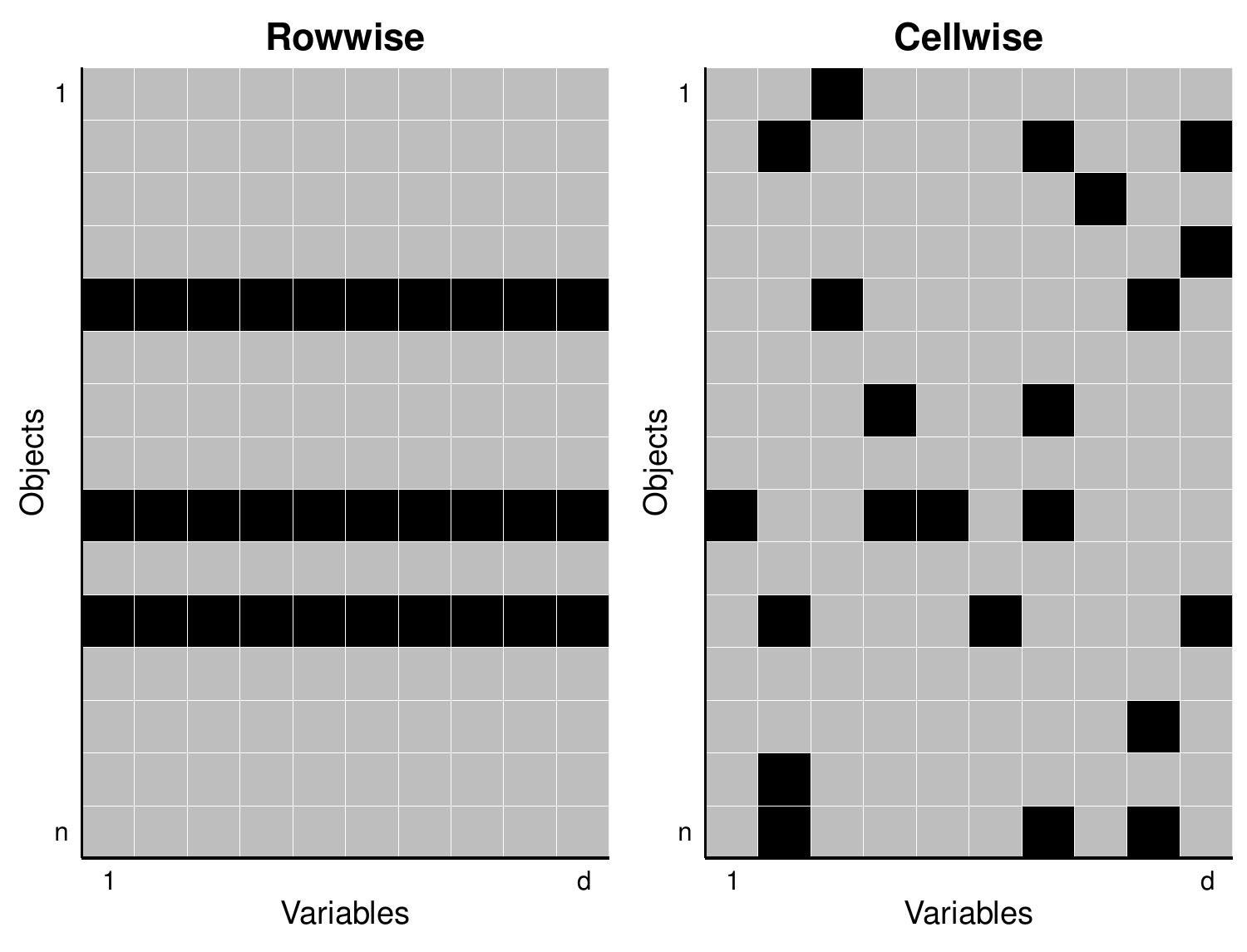}
\vspace{-0.3cm}
\caption{The rowwise and cellwise outlier paradigms
         (black means outlying).}
\label{fig:byrowbycell}
\end{figure}

In statistics and data analysis the word outlier typically 
refers to a row of the data matrix, as  
in the left panel of Figure~\ref{fig:byrowbycell}.
There has been much research since the 1960's 
to develop fitting methods that are less sensitive to such 
outlying rows, and that can detect the outliers by their
large residual (or distance) from that fit.
This topic goes by the name of robust statistics, see e.g.
the books by~\cite{Maronna:RobStat}
and~\cite{Rousseeuw:RobReg}.

Recently researchers have come to realize that the outlying
rows paradigm is no longer sufficient for modern 
high-dimensional data sets.
It often happens that most data cells (entries) in a 
row are regular and just a few of them are anomalous.
The first paper to formulate the cellwise paradigm 
was~\citep{Alqallaf:ICM}. They noted how outliers
propagate: given a fraction $\varepsilon$ of contaminated
cells at random positions, the expected fraction 
of contaminated rows is
\begin{equation}\label{eq:propagate}
  1 - (1 - \varepsilon)^d
\end{equation}
which quickly exceeds 50\% for increasing $\varepsilon$
and/or increasing dimension $d$,
as illustrated in the right panel of 
Figure~\ref{fig:byrowbycell}.
This is fatal because rowwise methods cannot handle more 
than 50\% of contaminated rows if one assumes some basic 
invariance properties, see e.g.~\citep{Lopuhaa:affine}.

The two paradigms are quite different.
The outlying row paradigm is about cases that do not belong
in the dataset, for instance because they are members of a
different population. 
Its basic units are the cases, and if you interpret them as
points in $d$-dimensional space they are indeed indivisible.
But if you consider a case as a row in a matrix then it can
be divided, into cells.
The cellwise paradigm assumes that some of these cells
deviate from the values they should have had, perhaps due 
to gross measurement errors, whereas the remaining cells
in the same row still contain useful information.

The anomalous data cells problem has proved to 
be quite hard, as the existing tools do not suffice. 
Recent progress was made by~\cite{Danilov:PhD},
~\cite{VanAelst:huberized},~\cite{Agostinelli:2SGS},
~\cite{Oellerer:ShootingS},
and~\cite{Leung:regression}.

\begin{figure}[!ht]
\centering
\includegraphics[width=7cm]{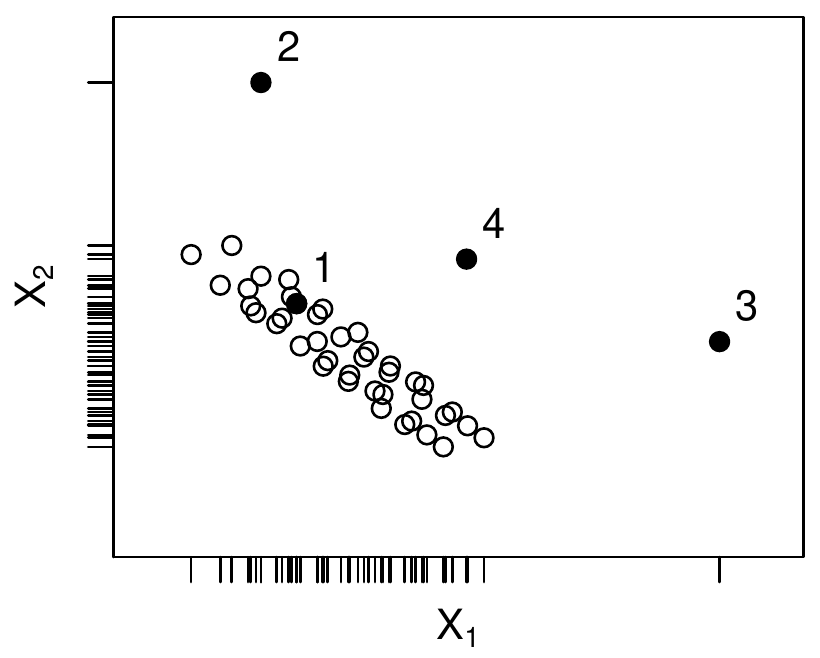}
\vspace{-0.3cm}
\caption{Illustration of bivariate outliers.}
\label{fig:toyexample}
\end{figure}

Note that in the right panel of Figure~\ref{fig:byrowbycell}
we do not know at the outset which (if any) cells are 
black (outlying), in stark contrast with the missing data
framework.
To illustrate the difficulty of detecting deviating data 
cells let us look at the artificial bivariate ($d=2$) 
example in Figure~\ref{fig:toyexample}.
Case 1 lies within the near-linear pattern of the
majority, whereas cases 2, 3, and 4 are outliers.
The first coordinate of case 2 lies among the majority
of the $x_{i1}$ (depicted as tickmarks under the 
horizontal axis) while its second coordinate $x_{22}$ is 
an outlying cell. 
For case 3, the first cell $x_{31}$ is outlying.
But what about case 4? Both of its coordinates fall
in the appropriate ranges.
Either coordinate of case 4 could be responsible for 
the point standing out, so the problem is not 
identifiable. 
We would have to flag the entire row 4 as outlying.

But now suppose that we obtain five more variables 
(columns) for these cases, which are correlated
with the columns we already have.
Then it may be possible to see that $x_{41}$ perfectly
agrees with the values in the new cells of row 4,
whereas $x_{42}$ does not.
Then we would conclude that the cell $x_{42}$ is 
the culprit. 
Therefore, this is one of the rare situations where 
a higher dimensionality is not a curse but can be turned 
into an advantage.
In fact, the method proposed in the next section typically 
performs better when the number of dimensions increases.

This example shows that deviating data cells do not need to 
stand out in their column: it suffices that they
disobey the pattern of relations between this variable 
and those correlated with it. 
Here we propose the first method for detecting deviating
data cells that takes the correlations between the variables 
into account.
Unlike existing methods, our {\it DetectDeviatingCells} (DDC) 
method is not restricted to data with over 50\% of clean rows.
Other advantages are that it can deal with high dimensions
and that it provides predicted values of the 
outlying cells.
As a byproduct it imputes missing values in the data.
For software availability see Section~\ref{sec:software}.

\section{\normalsize{BRIEF SKETCH OF THE METHOD}}
\label{sec:sketch}

The (possibly idealized) model states that the rows
$\bx_i$ were
generated from a multivariate Gaussian distribution with
unknown $d$-variate mean $\bmu$ and positive semidefinite
covariance matrix $\bSigma$, after which some cells were 
corrupted or became missing.
However, the algorithm will still run if the underlying
distribution is not multivariate Gaussian.

The method uses various devices from robust statistics and 
is described more fully in Section~\ref{sec:description}.
We only sketch the main ideas here.
The method starts by standardizing the data and flagging
the cells that stand out in their column. 
Next, each data cell is predicted based on the unflagged 
cells in the same row whose column is correlated with the 
column in question.
Finally, a cell for which the observed value differs much
from its predicted value is considered anomalous.
The method then produces an imputed data matrix, which
is like the original one except that cellwise outliers
and missing values are replaced by their predicted values.
The method can also flag an entire row if it contains 
too much anomalous behavior, so that we have the option of
downweighting or dropping that row in subsequent fits 
to the data.

This method looks unusual but it successfully navigates 
around the many pitfalls inherent in the problem, and 
worked the best out of the many approaches we tried.
Its main feature is the prediction of individual cells.
This can be characterized as a locally linear fit,  where 
`locally' is not meant in the usual sense (of Euclidean
distance) but instead refers to the space of variables 
endowed with a kind of correlation-based metric.
 
Along the way DDC 
imputes missing data values by their predicted values.
This is far less efficient than the EM algorithm 
of~\cite{Dempster:EM} when the data are Gaussian 
and outlier-free, but it is more robust against cellwise 
outliers.

The DDC method has the natural equivariance
properties. If we add a constant to all the values in a
column of $X$, or multiply any column by a nonzero factor,
or reorder the rows or the columns, the result
will change as expected.

\section{\normalsize{EXAMPLES}}
\label{sec:examples}
The first dataset was scraped from the website of the British 
television show Top Gear by~\cite{Alfons:robustHD}.
It contains 11 objectively measured numerical variables
about 297 cars. 
Five of these variables (such as Price) were rather 
skewed so we logarithmically transformed them.

\begin{figure}[!b]
\centering
\includegraphics[width=15cm,angle=0]
                {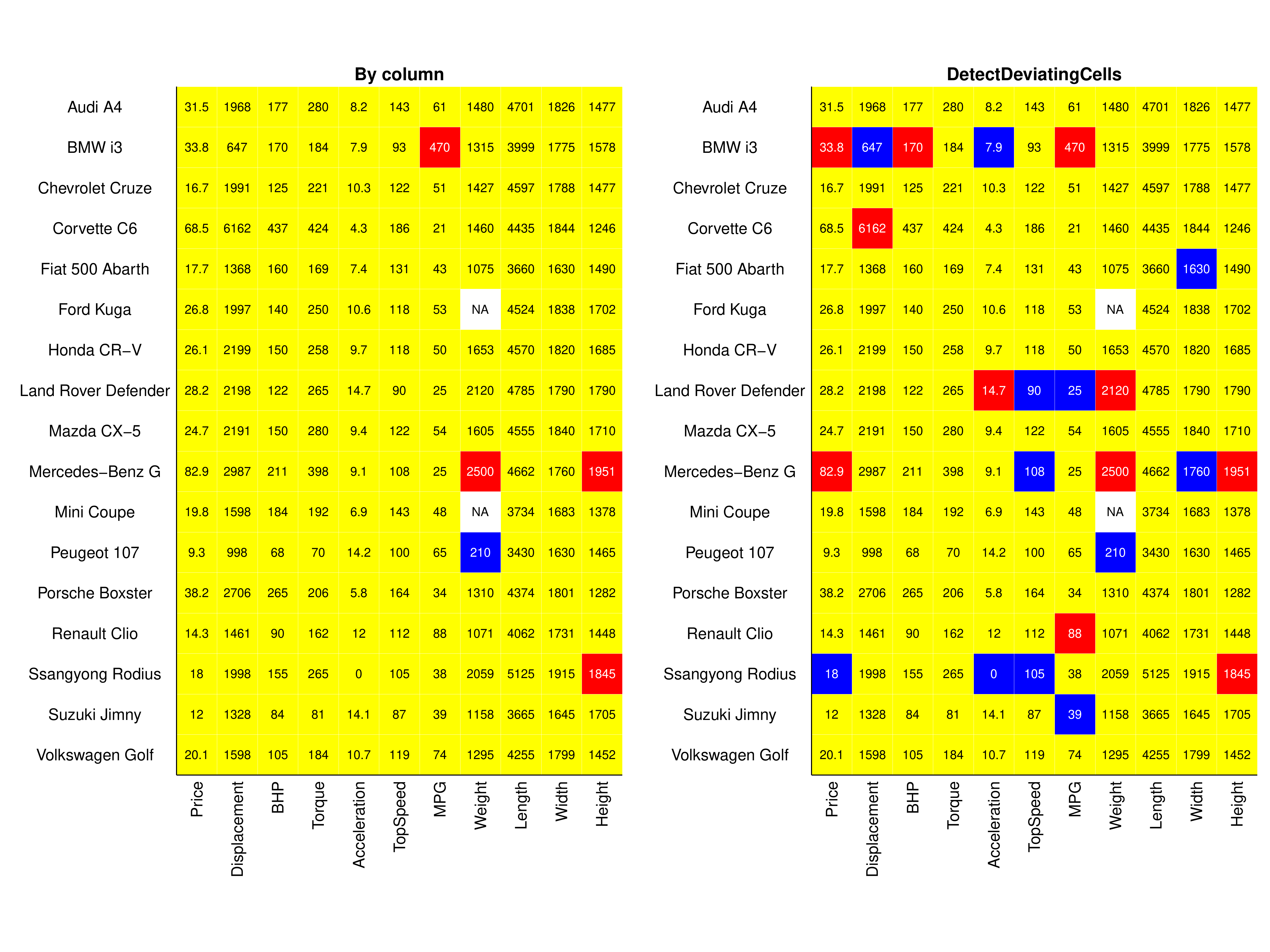}
\caption{Selected rows from cell maps of Top Gear data:
         (left) when detecting anomalous data cells per column; 
	       (right) when using the proposed method.}
\label{fig:TopGear}
\end{figure}

The left panel of Figure~\ref{fig:TopGear} shows the result 
of applying a simple univariate outlier identifier to each 
of the columns of this dataset separately. 
For column $j$ this method first computes
a robust location $m_j$ (such as the median) of its values,
as well as a robust scale $s_j$ and then computes the 
standardized values $z_{ij} = (x_{ij}-m_j)/s_j$.
A cell is then flagged as outlying if $|z_{ij}| > c$ where
the cutoff $c=2.576$ corresponds to a tolerance of 99\%
so that under ideal circumstances about 1\% of the cells would 
be flagged.
Most of the cells were yellow, indicating they were not
flagged. 
Here we only show some of the more interesting rows out of 
the 297.
Deviating cells whose $z_{ij}$ is positive are colored red,
and those with negative $z_{ij}$ are in blue.
Missing data are shown in white and labeled
NA (from `Not Available').

The right hand panel of Figure~\ref{fig:TopGear} shows the
corresponding cell map obtained by running 
{\it DetectDeviatingCells} on these data. 
In the algorithm we used the same 99\%
tolerance setting yielding the same cutoff $c=2.576$.
The color of the deviating cells now reflects whether the
observed cell value is much higher or much lower than the 
predicted cell value.

The new method shows much more structure than looking by
column only.
The high gas mileage (MPG) of the BMW i3 is no longer the only 
thing that stands out, and in fact it is an electric vehicle 
with a small additional gas engine.
The columnwise method does not flag the Corvette's displacement 
which is not that unusual by itself, but it is high in relation 
to the car's other characteristics.
None of the properties of the Land Rover Defender (an 
all-terrain vehicle) stand out on their own, but their 
combination does.
The weight of 210 kg listed for the Peugeot 107 is clearly 
an error, which gets picked up by both methods.
The univariate method only flags the height of the Ssangyong 
Rodius, whereas {\it DetectDeviatingCells} 
also notices that its acceleration time of zero seconds 
from 0 to 62 mph is too low compared to its other properties, 
and in fact it is physically impossible.

\begin{figure}
\centering
\includegraphics[height=16cm,angle=0]
                {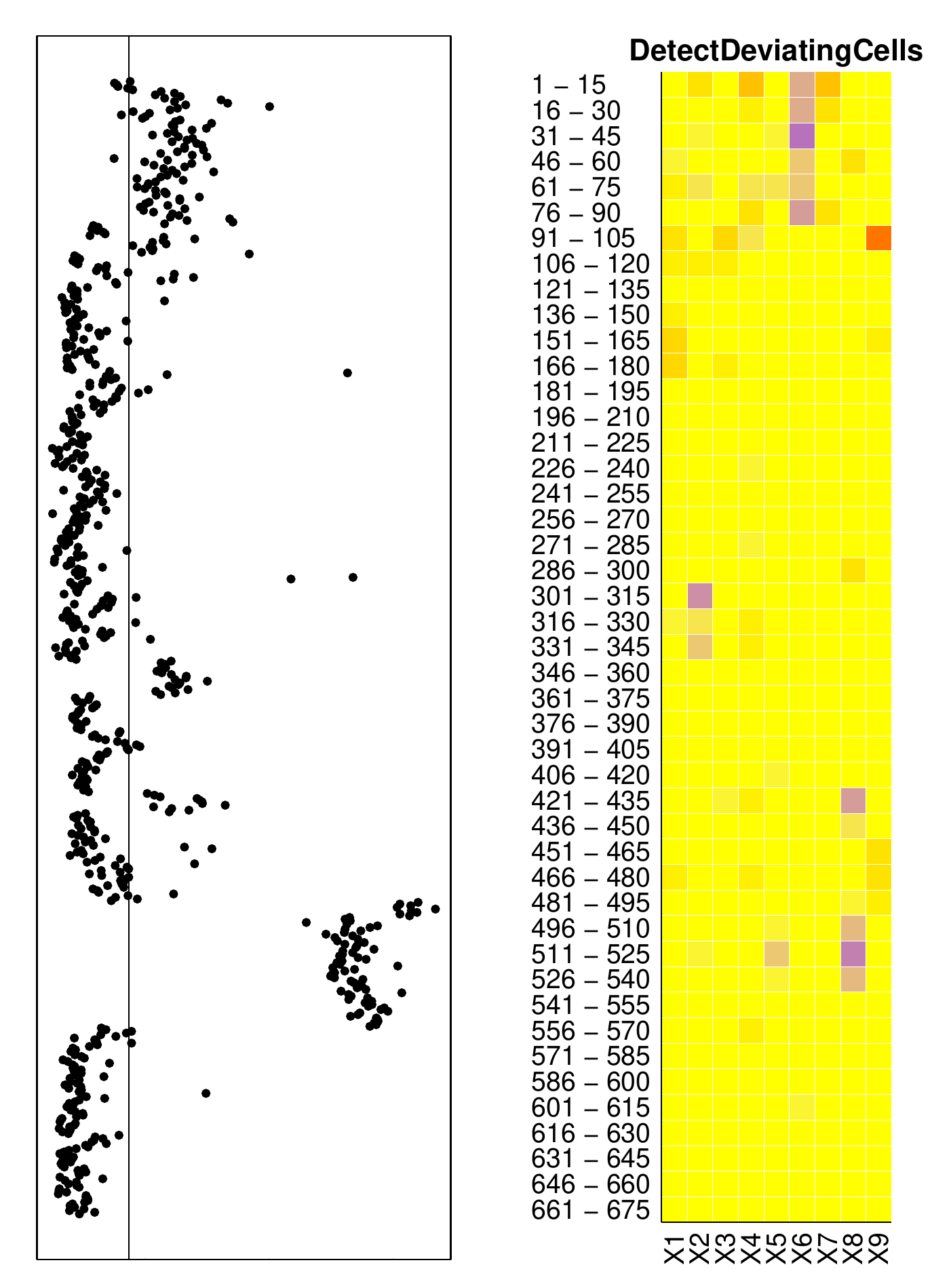}						
\caption{Left: robust distances of the rows
         of the Philips data.
				 Right: map of {\it DetectDeviatingCells} on
				 these data, combined per blocks of 15 rows.}
\label{fig:Philips}
\end{figure}

As a second example we take the Philips dataset which 
measures $d=9$ characteristics of $n=677$ TV parts
created by a new production line.
The left panel of Figure~\ref{fig:Philips} is a rotated 
version of Figure 9(a) in~\citep{Rousseeuw:FastMCD} and 
shows, from top to bottom, the robust distance $RD_i$ for
$i=1,\ldots,n$. These were computed as
\begin{equation*}
\mbox{RD}_i = \sqrt{(\bx_i-\bT)'\bS^{-1}(\bx_i-\bT)}
\end{equation*}
where $\bT$ and $\bS$ are the estimates of location and 
scatter obtained by the Minimum Covariance Determinant 
(MCD) from~\citep{Rousseeuw:HBD}, a rowwise robust method.
It shows that the process was out of control in the
beginning (roughly rows 1--90) and near the end
(around rows 480--550).
However, by itself this does not yet tell us what 
happened in those products.

In the right panel we see the cell map of DDC. 
The analysis was carried out on all 677 rows, but in
order to display the results we created blocks of
15 rows and 1 column. The color of each cell
is the `average color' over the block.
The resulting cell map indicates that in the 
beginning variable 6 had lower values than would be
predicted from the remaining variables, whereas the 
later outliers were linked with lower values of 
variable 8.
(In between, the cell map gives hints about some more 
isolated deviations.)
We do not claim that this is the complete answer, but at 
least it gives an initial indication of what to look
for.
(DDC also flagged some rows, but this is not shown in the 
cell map to avoid confusion.)
Note that these data are not literally a multivariate
sample because the rows were provided consecutively,
but both the MCD and DDC methods ignore the time order of 
the rows.

\begin{figure}[bt]
\centering
\includegraphics[width=0.91\textwidth]
                {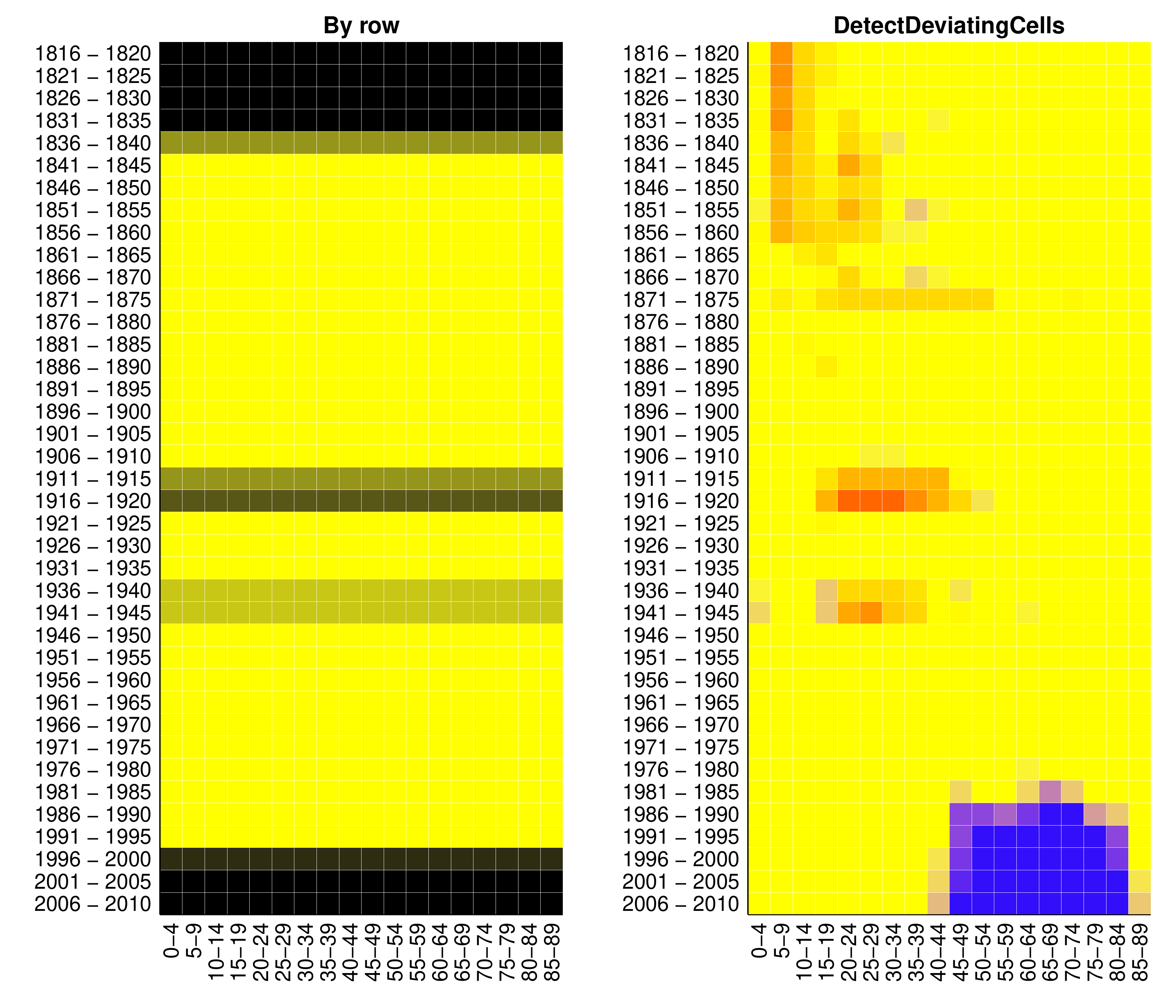}
\caption{Male mortality in France in 1816--2010: 
         (left) detecting outlying
         rows by a robust PCA method; (right) detecting
         outlying cells by {\it DetectDeviatingCells}.
	 After the analysis, the cells were grouped in
	 blocks of $5 \times 5$ for visibility.}
\label{fig:mortality}
\end{figure}

The next example is the mortality by age for males in 
France, from 1816 to 2010 as obtained from
the~\cite{HMD:HMD}. An earlier version was analyzed 
by~\cite{Hyndman:boxplot}. 
Each case corresponds to the mortalities in a given year.
The left panel in Figure~\ref{fig:mortality} was obtained 
by ROBPCA~\citep{Hubert:ROBPCA}, a rowwise robust method
for principal component analysis.
Such a method can only detect entire rows, and the rows it 
has flagged are represented as black horizontal lines.
A black row doesn't reveal any information about its cells.
The analysis was carried out on the data set with the
individual years and the individual ages, but as this
resolution would be too high to fit on the page we have
combined the cells into $5 \times 5$ blocks afterward.
The combination of some black rows with some yellow ones 
has led to gray blocks.
We can see that there were outlying rows in the early
years, the most recent years, and during two periods
in between.

By contrast, the right panel in Figure~\ref{fig:mortality}
identifies a lot more information.
The outlying early years saw a high infant mortality.
During the Prussian war and both world wars there was a 
higher mortality among young adult men.
And in recent years mortality among middle-aged and older 
men has decreased substantially, perhaps due to medical 
advances.

\begin{figure}[h]
\centering
\includegraphics[width=15.2cm,angle=0]
                {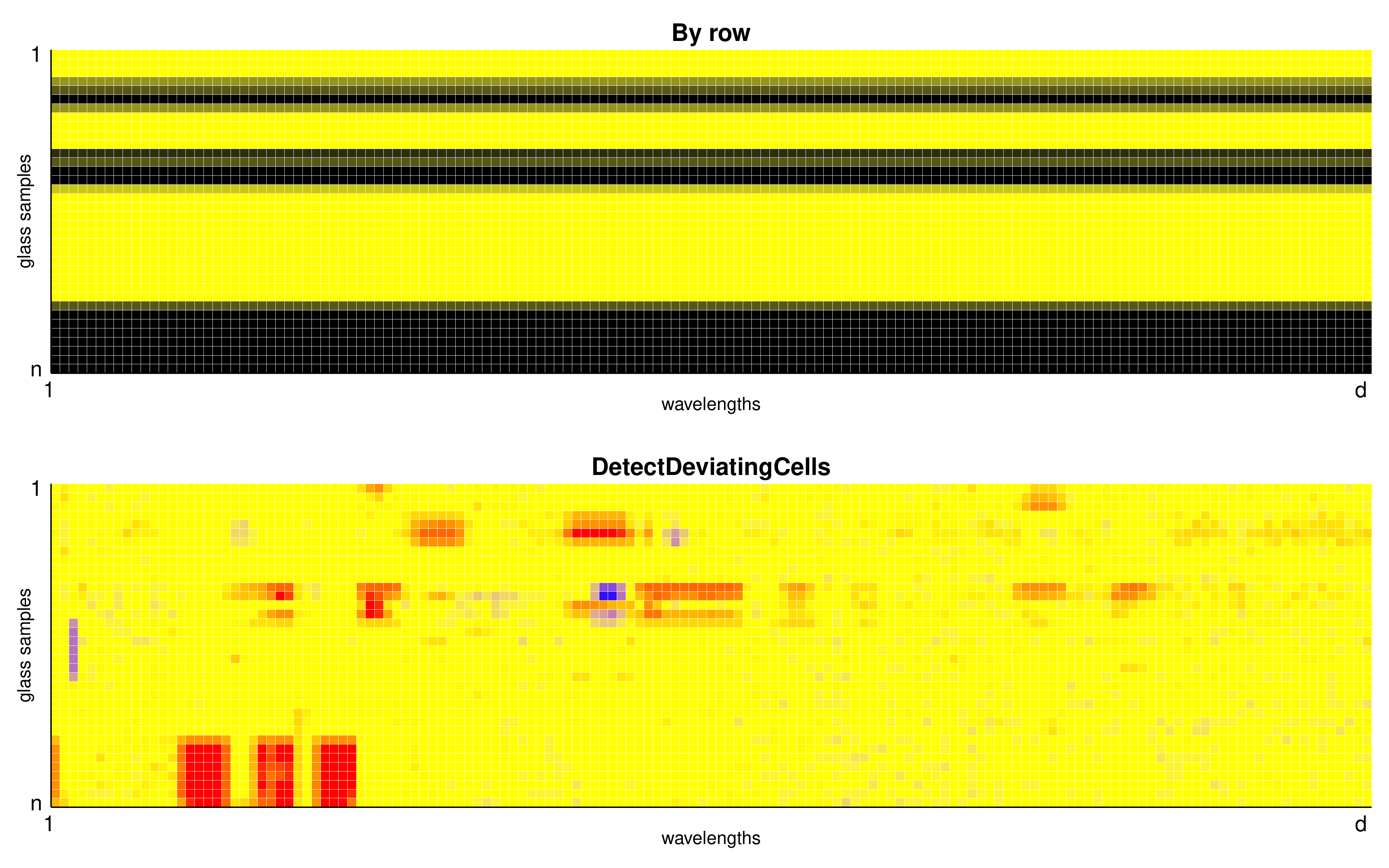}								
\caption{Cell maps of $n=180$ archeological glass
   spectra with $d=750$ wavelengths. 
	 The positions of the deviating data cells reveal 
	 the chemical contaminants.}
\label{fig:glass}
\end{figure}

Our final example consists of spectra with $d = 750$
wavelengths collected on $n = 180$ archeological glass 
samples~\citep{Lemberge:glass}.
Here the number of dimensions (columns) exceeds the 
number of cases (rows).
{\it DetectDeviatingCells} has no problems with this, 
and the R implementation took about 1 minute on a laptop 
with Intel i7-4800MQ 2.7GHz processor for this dataset.
The top panel in Figure~\ref{fig:glass} shows the rows 
detected by the robust principal components method.
The lower panel is the cell map obtained by 
{\it DetectDeviatingCells} on this $180 \times 750$ 
dataset.
After the analysis, the cells were again grouped in 
$5 \times 5$ blocks.
We now see clearly which parts of each spectrum are
higher/lower than predicted.
The wavelengths of these deviating cells reveal the
chemical elements responsible.

In general, the result of {\it DetectDeviatingCells} can 
be used in several ways:
\begin{enumerate}
\item Ideally, the user 
      looks at the deviating cells and whether their
      values are higher or lower than predicted, and 
      makes sense of what is going on.
      This may lead to a better understanding of the
      data pattern, to changes in the way the data
      are collected/measured, to dropping certain
      rows or columns, to transforming variables,
      to changing the model, and so on.
\item If the data set is too large for visual inspection
      of the results or the analysis is
      automated, the deviating cells can be set to missing
      after which the data set is analyzed by a method
      appropriate for incomplete data.
\item If no such method is available, one can analyze
      the imputed data set $\bX_{imp}$ 
      produced by {\it DetectDeviatingCells} which has no
      missings. 
\end{enumerate}
In 2. and 3. one has the option to drop the flagged 
rows before taking the step.
If that step is carried out by a sparse method 
such as the Lasso~\citep{Tibshirani:Lasso,Staedler:PA}
or another form of variable selection, one would of 
course look more closely at the deviating cells in the 
variables that were selected.

{\bf Remark.} In the above examples the cellwise
outliers happened to be mainly concentrated in fewer than 
half of the rows, allowing the rowwise 
robust methods MCD and ROBPCA to detect most of those rows.
This may give the false impression that it would suffice to 
apply a rowwise robust method and then to analyze the 
flagged rows further.
But this is not true in general: by~\eqref{eq:propagate} 
there will typically be too many contaminated rows, so 
rowwise robust methods will fail.

\section{\normalsize{DETAILED DESCRIPTION OF THE ALGORITHM}}
\label{sec:description}

The {\it DetectDeviatingCells} (DDC) algorithm has been
implemented in R and Matlab (see Section~\ref{sec:software}).
As described before, the underlying model is that the data
are generated from a multivariate Gaussian distribution
$N(\bmu,\bSigma)$ but afterward some cells were corrupted
or omitted.
The variables (columns) of the data should thus be 
numerical and take on more than a few values.
Therefore the code does some preprocessing: it identifies
the columns that do not satisfy this condition, such 
as binary dummy variables, and performs its computations 
on the remaining columns.

These remaining variables should then be approximately 
Gaussian in their center, that is, apart from any outlying 
values.
This could be verified by QQ plots, and one could flag
highly skewed or long-tailed variables. 
It is recommended to transform very non-gaussian variables
to approximate gaussianity, like we took the logarithm 
of the right-skewed variable Price in the cars data in
Section 3.
More general tools are the Box-Cox and Yeo-Johnson
\citep{Yeo:transform} 
transformations, that have parameters which need 
to be estimated as done by \cite{Bini:transformation} and 
\cite{Riani:transformation}. 
We chose not to automate this step because we feel it is 
best carried out by a person with subject matter knowledge
about the data.
From here onward we will assume that the variables are
approximately Gaussian in their center.
The algorithm does not require joint normality of variables
in order to run. 

The actual algorithm then proceeds in eight steps.

{\bf Step 1: standardization.} For each column $j$ of 
$\bX$ we estimate 
\begin{equation}\label{eq:locsca}
  m_j = robLoc_i(x_{ij}) \quad \mbox{  and  } \quad
	s_j = robScale_i(x_{ij} - m_j)
\end{equation}
where {\it robLoc} is a robust estimator of location,
and {\it robScale} is a robust
estimator of scale which assumes its argument has already 
been centered.
The actual definitions of these estimators are given in
the Appendix. 
Next, we standardize $\bX$ to $\bZ$ by
\begin{equation}\label{eq:standardize} 
  z_{ij} = (x_{ij}-m_j)/s_j \;\;.
\end{equation}

{\bf Step 2: apply univariate outlier detection to all
 variables.}
After the columnwise standardization 
in~\eqref{eq:standardize} we define a new matrix $\bU$ 
with entries
\begin{equation}\label{eq:unidetect}
  u_{ij} =
\left\{
\begin{array}{ll}
  z_{ij} & \text{ if } |z_{ij}| \leqslant c \\
  \text{NA} & \text{ if } |z_{ij}| > c \;\;. \\
\end{array}
\right. 
\end{equation}
Due to the standardization in~\eqref{eq:standardize}, 
formula~\eqref{eq:unidetect} is a columnwise outlier 
detector.
The cutoff value $c$ is taken as
\begin{equation}\label{eq:qCut}
       c = \sqrt{\chi^2_{1,p}}
\end{equation}
where $\chi^2_{1,p}$ is the $p$-th quantile of the
chi-squared distribution with 1 degree of freedom,
where the probability $p$ is 99\% by default so 
that under ideal circumstances only 1\% of the entries 
gets flagged.

{\bf Step 3: bivariate relations.} 
For any two variables $h \neq j$ we compute
their correlation as
\begin{equation}\label{eq:cor}
   \mbox{cor}_{jh} = robCorr_i(u_{ij},u_{ih})
\end{equation}
where {\it robCorr} is a robust correlation measure
given in the Appendix (the computation is over all $i$
for which neither $u_{ij}$ or $u_{ih}$ is NA).
From here onward we will only use the relation between
variables $j$ and $h$ when
\begin{equation}\label{eq:corrlim}
  |\mbox{cor}_{jh}| \geqslant corrlim
\end{equation}
in which {\it corrlim} is set to 0.5 by default.
Variables $j$ that satisfy \eqref{eq:corrlim} for some
$h \neq j$ will be called {\it connected}, and contain
useful information about each other.
The others are called {\it standalone} variables.
For the pairs $(j,h)$ satisfying~\eqref{eq:corrlim}
we also compute 
\begin{equation}\label{eq:b}
   b_{jh} = robSlope_i(u_{ij} | u_{ih})
\end{equation}
where {\it robSlope} computes the slope of a robust
regression line without intercept that predicts variable 
$j$ from variable $h$ (see Appendix).
These slopes will be used in the next step to provide
predictions for connected variables.

{\bf Step 4: predicted values.} Next we compute predicted 
values $\hat{z}_{ij}$ for all cells. 
For each variable $j$ we consider the set
$H_j$ consisting of all variables $h$ satisfying
\eqref{eq:corrlim}, including $j$ itself.
For all $i=1,\ldots,n$ we then set
\begin{equation}\label{eq:combination}
  \hat{z}_{ij} = G(\{b_{jh}u_{ih}\;; 
	               \;h \mbox{ in}\;\; H_j\})
\end{equation}
where $G$ is a combination rule applied to these numbers,
which omits the NA values and is zero when no values remain.
(The latter corresponds to predicting the unstandardized
cell by the estimated location of its column, which is a
 fail-safe.)
Our current preference for $G$ is a weighted mean with
weights $w_{jh} = |cor_{jh}|$ but other choices are possible,
such as a weighted median.
The advantage of \eqref{eq:combination} is that the
contribution of an outlying cell $z_{ih}$ to $\hat{z}_{ij}$ 
is limited because $|u_{ih}| \leqslant c$ and it can only 
affect a single term, which would not be true for a
prediction from a least squares multiple regression of 
$z_{.j}$ on the $d-1$ remaining variables $z_{.h}$ together.
We will illustrate the accuracy of the predicted cells
by simulation in Section~\ref{sec:comparison}.
	
{\bf Step 5: deshrinkage.} 
Note that a prediction such as \eqref{eq:combination} tends 
to shrink the scale of the entries, which is undesirable.
We could try to shrink less in the individual terms
$b_{jh}u_{ih}$ but this would not suffice because
these terms can have different signs for different $h$. 
Therefore, we propose to deal with the shrinkage {\it after} 
applying the combination rule \eqref{eq:combination}.
For this purpose we replace $\hat{z}_{ij}$ by
$a_j \hat{z}_{ij}$ for all $i$ and $j$, where
\begin{equation}
   a_j := robSlope_{i'}(z_{i'j}|\hat{z}_{i'j})
\end{equation}
comes from regressing the observed $z_{.j}$ on the
(shrunk) predicted $\hat{z}_{.j}$\;.

{\bf Step 6: flagging cellwise outliers.}
In steps 4 and 5 we have computed predicted values
$\hat{z}_{ij}$ for all cells. 
Next we compute the standardized cell residuals
\begin{equation}\label{eq:stdres}
   r_{ij} = \frac{z_{ij} - \hat{z}_{ij}}
	  {robScale_{i'}(z_{i'j} - \hat{z}_{i'j})} \;\;.
\end{equation}
In each column $j$ we then flag all cells with 
$|r_{ij}| > c$ as anomalous, where $c$ was given
in~\eqref{eq:qCut}.
We also assemble the `imputed' matrix $\bZ_{imp}$ which
equals $\bZ$ except that it replaces deviating cells
$z_{ij}$ and NA's by their predicted values $\hat{z}_{ij}$. 
The unflagged cells remain as they were.

We have the option of setting the
flagged cells to NA and repeating steps 4 to 6 in
order to improve the accuracy of the estimates.
This can be iterated.

{\bf Step 7: flagging rowwise outliers.}
In order to decide whether to flag row $i$ we could just 
count the number of cells whose $|r_{ij}|$ exceeds a
cutoff $a$, but this would miss rows with many fairly large
$|r_{ij}|\;<\;a$ .
The other extreme would be to compare $\ave_j(r_{ij}^2)$ 
to a cutoff, but then a row with one very outlying cell 
would be flagged as outlying, which would defeat the 
purpose.
We do something in between. 
Under the null hypothesis of multivariate Gaussian data 
without any outliers, the distribution of the $r_{ij}$ 
is close to standard Gaussian, so the cdf of $r_{ij}^2$
is approximately the cdf $F$ of $\chi^2_1$. 
This leads us to the criterion
\begin{equation}\label{eq:rowcrit}
   T_i = \ave_{j=1}^d F(r_{ij}^2)\;\;.
\end{equation} 
We then standardize the $T_i$ as in~\eqref{eq:standardize} 
and flag the rows $i$ for which the standardized $T_i$
exceed the cutoff $c$ of~\eqref{eq:qCut}.

When we find that row $i$ has an unusually
large $T_i$ this does not necessarily `prove' that it
corresponds to a member of a different population, but 
at least it is worth looking into.

Even though the $T_i$ can flag {\it some} 
rowwise outliers, there are types of rowwise outliers that 
it may not detect, for instance in the barrow wheel 
configuration of \cite{Stahel:wheel} where a rowwise
outlier may not have any large $r_{ij}^2$. 
Therefore, it is recommended to use rowwise robust methods 
in subsequent analyses of the data, after dropping the
rows that were flagged.

{\bf Step 8: destandardize.}	
Next, we turn the imputed matrix $\bZ_{imp}$ into an 
imputed matrix $\bX_{imp}$ by undoing the standardization
in \eqref{eq:standardize}.
The main output of DDC is 
$\bX_{imp}$ together with the list of cells flagged as
outlying and, optionally, the list of rowwise outliers
that were found.\\ 					

Predicting the value of a cell in steps 4
and 5 above evokes the imputation of a missing value, 
as in the EM algorithm of~\cite{Dempster:EM}. 
However, there are two reasons why the EM method may
not work in the present situation.
First of all, we would need estimators of location and
scatter $\hat{\bmu}$ and $\hat{\bSigma}$ 
that are robust against cellwise and rowwise 
outliers, which we only know how to obtain {\it after} 
detecting those outliers. 
And secondly, the conditional expectation of $x_{ij}$ is 
a linear 
combination of all the available $x_{ih}$ with $h \neq j$ 
but some of those cells may be outlying themselves,
thus spoiling the sum.
 
The DDC method employs measures of bivariate correlation, 
and regression for bivariate data.
This may seem naive compared to estimating scatter matrices 
and/or performing multiple regression on sets of $q > 2$ 
variables.
However, the latter would imply that we must compute 
scatter matrices between $q$ variables in an environment 
where a fraction $\varepsilon$ of cells is contaminated, 
so that according to~\eqref{eq:propagate}
a fraction $1 - (1 - \varepsilon)^q$ of the rows 
of length $q$ is expected to be contaminated.
The latter fraction grows very quickly with $q$,
and once it exceeds 50\% there is no hope of estimating the
scatter matrix by a rowwise robust method.
So the larger $q$ is, the less this approach can work.
Therefore we chose the smallest possible value $q = 2$,
which has the additional advantage that the computational 
effort to robustly estimate bivariate scatter is much lower
than for higher $q$.

To illustrate this, Table 1 shows the effect of fitting
substructures in dimension $q < d$ for various values 
of $q$.
The total time complexity is the product of the time
needed for a 50\%-rowwise-breakdown fit (regression 
or scatter) to a $q$-variate data set (in which 
elemental sets are formed by $q-1$ data points and the
 origin) and
the number of combinations of $q$ out of $d$ variables
(for large $d$). 
The complexity grows very fast with $q$.
The next column is the expected breakdown value, i.e. 
what fraction of evenly distributed cellwise outliers 
these fits can typically withstand, i.e. $1 - 2^{-1/q}$.
The expected breakdown value drops quickly with $q$. 
The final column is the probability that a subrow
of $q$ entries contains no outliers when the probability
of a cell being outlying is 10\%.
Table 1 is in fact overly optimistic
because it does not account for the problem that the
predictions from such $q$-variate fits are affected
whenever at least one of the cells in it is outlying. 

\begin{table}
\begin{center}
\caption{Fitting substructures in $q < d$ dimensions}
\vspace{0.3cm}
\begin{tabular}{ccrc}
\hline
\vspace{-0.3cm}
       & total time & breakdown & probability of \\
	$q$  & complexity & value     & clean $q$-row$^*$ \\
\hline 
\hline
 2   & $n\,d\,log(d)$  & 29.3\%  & 81.0\% \\
 3   & $n^2 d^3$       & 20.6\%  & 72.9\% \\
 4   & $n^3 d^4$       & 15.9\%  & 65.6\% \\
 5   & $n^4 d^5$       & 12.9\%  & 59.0\% \\
 10  & $n^9 d^{10}$    & 6.7\%   & 34.9\% \\
 20  & $n^{19} d^{20}$ & 3.4\%   & 12.2\% \\
\hline
\end{tabular}\\
\footnotesize{$^*$ Assuming a 10\% probability of a cell being 
       outlying.} 
\end{center}
\end{table}

This should not be confused with the approach of
\cite{Kriegel:SOD} who look for rowwise outliers 
under the assumption that outlyingness of a row is due
to only a few variables whereas the other variables are 
deemed irrelevant.
That situation is different from both the general rowwise
outlier setting and the cellwise outlier model, in each 
of which all variables may be relevant.

As described, the computation time (number of operations) 
of DDC is on the order of $O(nd^2)$ due 
to computing all $d(d-1)/2$ correlations between 
variables, where each correlation takes $O(n)$ time. 
The required memory (storage) space is then also $O(nd^2)$.
However, if the dimension $d$ is over (say) 1000 we  
switch to predicting each column by the $k$
columns that are most correlated with it, thereby bringing 
down the space requirement to $O(nd)$ which cannot be
reduced further as it is proportional to the size of the 
data matrix itself.
On a parallel architecture we can reduce the
computation time by letting each processor compute the
$k$ correlations for a subset of columns to be predicted.
On a non-parallel system we could switch to a fast
approximate $k$-nearest neighbor method such as the 
celebrated algorithm of~\cite{Arya:ANN} 
which would lower the computation time from
$O(nd^2)$ to $O(nd\,log(d))$.
Compared to analyzing each column separately, this algorithm
only spends an additional time factor $log(d)$ which grows 
very slowly with $d$.
A related speedup is used in Google 
Correlate~\citep{Vanderkam:Google}.

\section{\normalsize{COMPARISON TO OTHER DETECTION METHODS}}
\label{sec:comparison}
We now compare DDC to a method that 
detects outlying cells per column, the univariate GY filter 
of~\cite{Gervini:filter} as described in\linebreak 
\citep{Agostinelli:2SGS}. 
Applying the univariate GY filter to the columns of the cars 
data actually gives the same result as the simpler columnwise detection in Figure~\ref{fig:TopGear}.
The key difference between these methods is that the GY
filter uses an adaptive cutoff rather than the fixed
cutoff \eqref{eq:qCut}.

We will only report a small part of our simulations, the 
results of other settings being similar.
First, clean multivariate Gaussian data were generated with 
$\bmu = \bzero$ and two types of covariance matrices $\bSigma$
with unit diagonal. 
The ALYZ~\citep{Agostinelli:2SGS} random correlation matrices 
yield relatively low correlations between the variables, 
whereas the A09 correlation matrix given by 
$\rho_{jh} = (-0.9)^{|h-j|}$ yields both 
high and low correlations.

Next, these clean data were contaminated.
Outlying {\it cells} were generated by replacing a random
subset (say, 10\%) of the $n \times d$ cells by a value 
$\gamma$ which was varied to see its effect.

Outlying {\it rows} were generated in the hardest direction, 
that of the last eigenvector $\bv$ of the true
covariance matrix $\bSigma$. 
Next we rescale $\bv$ to the typical size of a data point, 
by making $\;\bv'{\bSigma}^{-1}\bv = E[Y^2] = d\;$ 
where $\;Y^2 \sim \chi^2_d\;$.
We then replaced a random set of rows of $\bX$ 
by $\gamma \bv$.

\vskip0.2cm
\begin{figure}[!ht]
\centering
\includegraphics[width=11.5cm,angle=0]
                {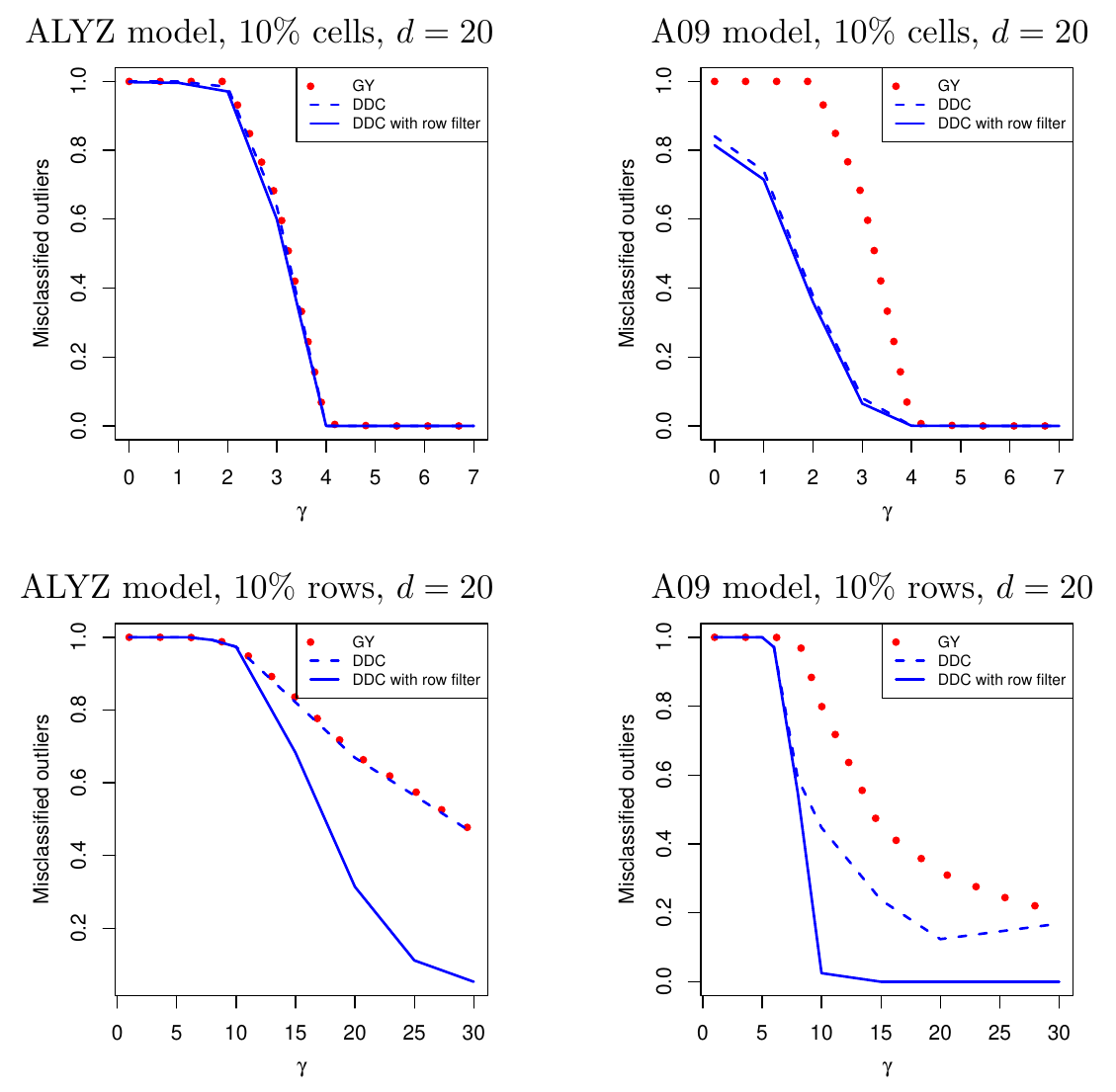}	
\vspace{-0.4cm}
\caption{Outlier misclassification rates of different detection 
         methods: (top) under cellwise contamination, (bottom)
         under rowwise contamination.
				 Here GY is the Gervini-Yohai filter.}
\label{fig:misclass_GY_DDC}
\end{figure}

Figure~\ref{fig:misclass_GY_DDC} shows the outlier
misclassification rate (for $n=200$ 
and $d=20$).
This is the number of cells that were generated as
outlying and detected by the method, divided by the 
total number of outlying cells generated.
This fraction depends on $\gamma$: it is high for small 
$\gamma$ (but those cells 
or rows are not really outlying) and then goes down.
In the top panels we see that all methods for detecting cells 
work about equally well when the correlations between variables 
are fairly small, whereas the new method does better when there 
are at least some higher correlations.
The bottom panels show a similar effect for detecting rows, 
and indicate that using the row filter, i.e. step 7 in 
Section~\ref{sec:description}, 
is an important component of DDC. 

\vskip0.2cm
\begin{figure}[!ht]
\centering
\includegraphics[width=11.5cm,angle=0]
                {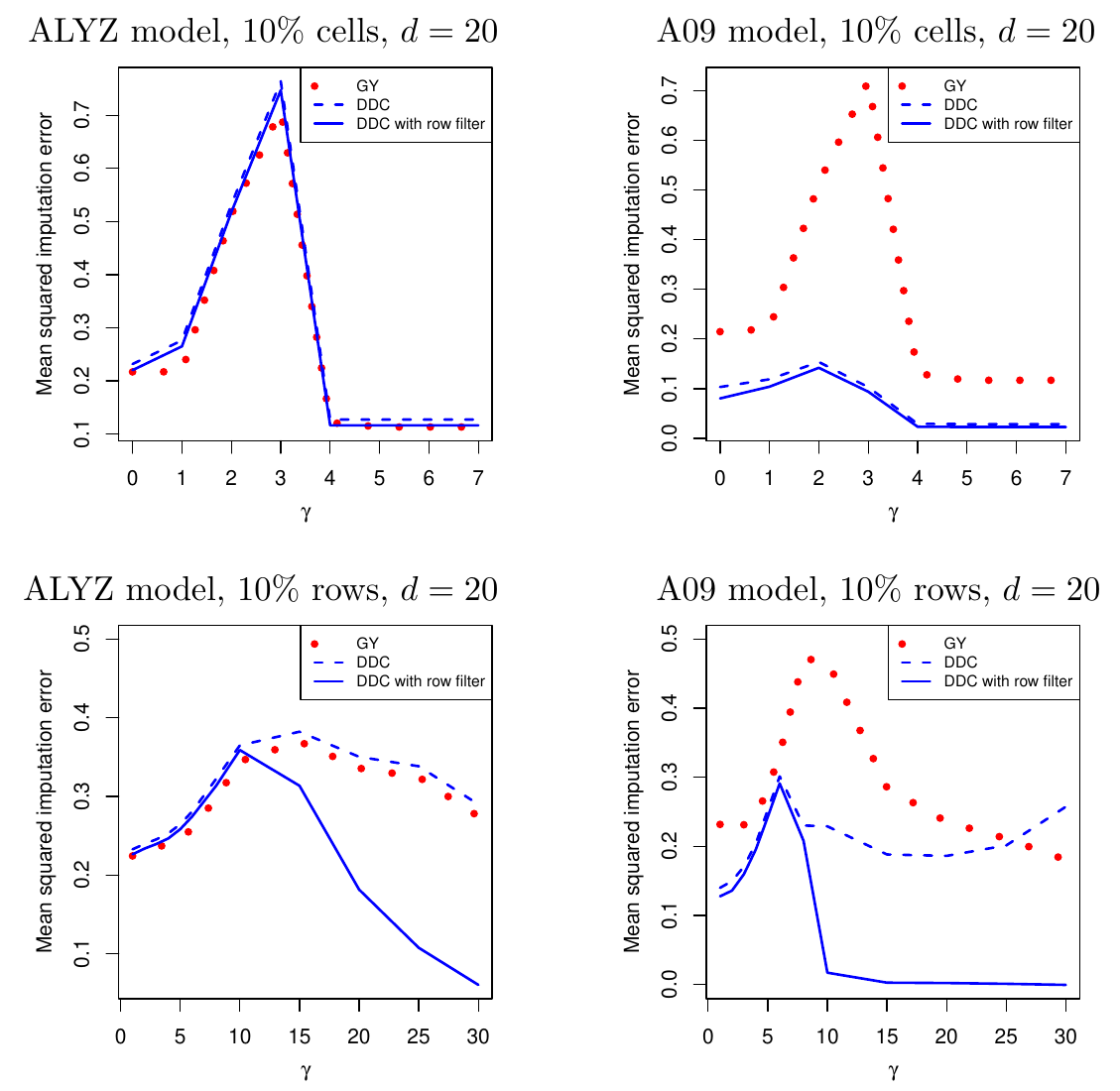}		
\vspace{-0.4cm}
\caption{Mean squared error of the imputed cells in
         the same simulation.} 
\label{fig:MSE_GY_DDC}
\end{figure}

The same simulation yields Figure~\ref{fig:MSE_GY_DDC}, 
which shows the mean squared error of the imputed cells
relative to the original cell values (before any
contamination took place), in which flagged rows were
taken out. 
Again DDC does best when there are some higher 
correlations.
This confirms the usefulness of computing predicted
cells in step 4 of the DDC algorithm.

We have repeated the simulations leading to 
Figures~\ref{fig:misclass_GY_DDC}
and~\ref{fig:MSE_GY_DDC} for dimensions $d$ ranging
upto 200, $n$ between 10 and 2000, and contamination
fractions upto 20\%, each time with 50 replications.
The method does not suffer from high dimensions, and
even for $n=20$ cases in $d=200$ dimensions
the comparisons between these methods yield 
qualitatively the same conclusions. 

\section{\normalsize{TWO-STEP METHODS FOR LOCATION AND SCATTER}} 
\label{sec:twostep}

Now assume that we are interested in estimating the
parameters of the multivariate Gaussian distribution
$N(\bmu,\bSigma)$ from which the data were generated 
(before cells and/or rows were contaminated). 

If the data were clean one would estimate $\bmu$ and 
$\bSigma$ by the mean of the data and the classical 
covariance matrix.
However, when the data may contain both cellwise and rowwise 
outliers the problem becomes much more difficult.
For this~\cite{Agostinelli:2SGS} proposed a two-step 
procedure called 2SGS which is the current best method.
The first step applies the univariate GY filter (from the
previous section) to each column of the data matrix $\bX$
and sets the flagged cells to NA.
The second step then applies the sophisticated Generalized 
S-Estimator 
(GSE) of~\cite{Danilov:GSE} to this incomplete data set, 
yielding $\bhmu$  and $\bhSigma$\;.
The GSE is a rowwise robust estimator 
of $\bmu$ and $\bSigma$ 
that was designed to work with data containing missing values,
following earlier work by~\cite{Little:robust}
and~\cite{Cheng:missing}. 

Our version of this is to replace GY in the first step 
by DDC, followed by the same second step.
When the first step flags a row we take it out of the
subsequent computations.
We also include the Huberized Stahel-Donoho (HSD) estimator 
of~\cite{VanAelst:huberized}, as well as the
Minimum Covariance Determinant (MCD)
estimator~\citep{Rousseeuw:HBD,Rousseeuw:FastMCD} which is 
robust to rowwise outliers but not to cellwise outliers.
For each method we measure how far $\bhSigma$ is from the 
true $\bSigma$ by the likelihood ratio type deviation
$$\mbox{LRT} = \mbox{trace}(\bhSigma \bSigma^{-1}) - 
     \mbox{log}(\mbox{det}(\bhSigma \bSigma^{-1})) - d$$
(which is a special case of the Kullback-Leibler divergence)
and average this quantity over all replications.

\begin{figure}[!htb]
\centering
\includegraphics[width=11.5cm,angle=0]
                {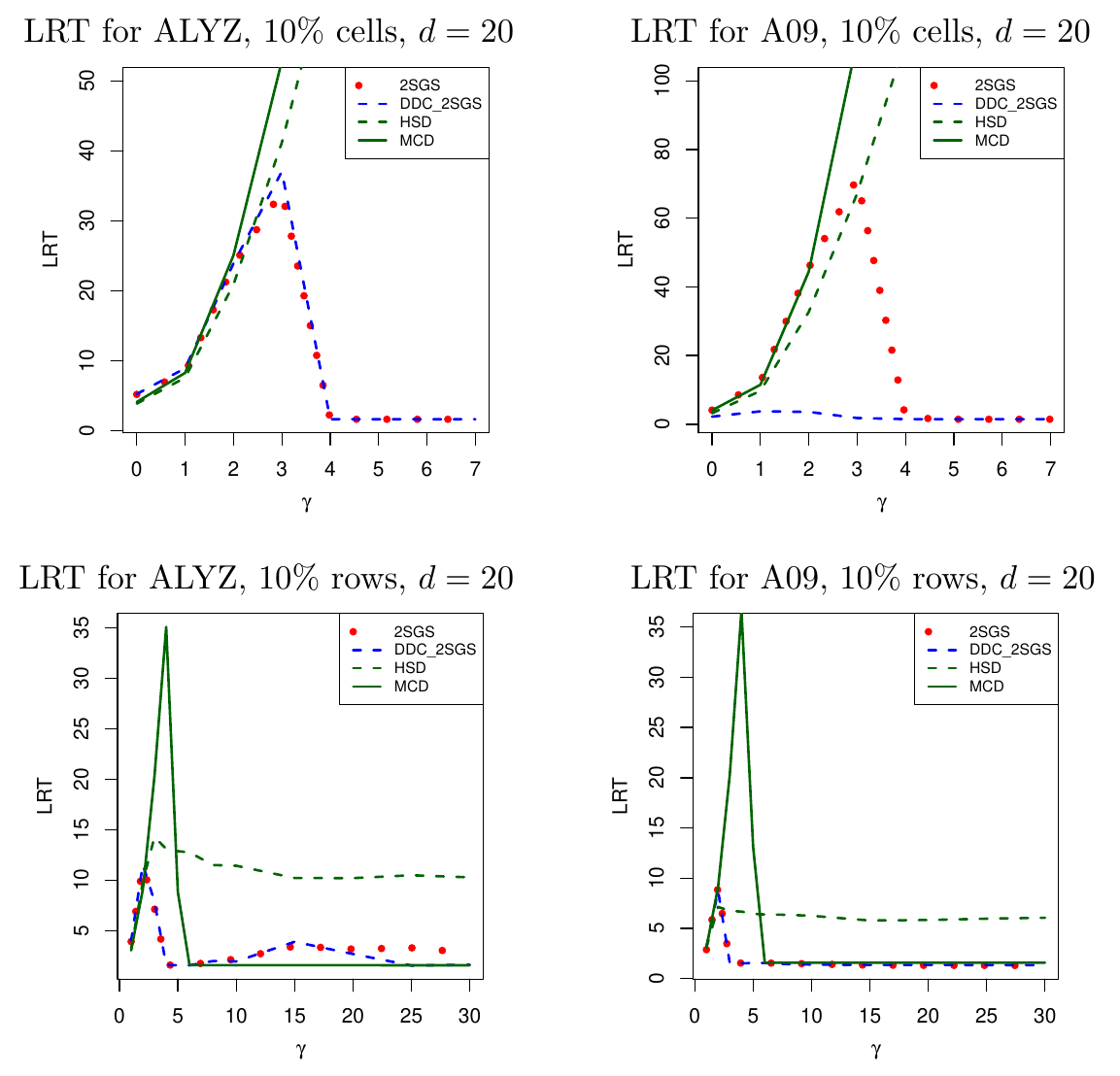}	
\vspace{-0.4cm}
\caption{LRT deviation of three estimates
         from the true scatter matrix.}
\label{fig:LRT_2SGS}
\end{figure}

Figure~\ref{fig:LRT_2SGS} compares these methods in the 
same simulation settings as Figures~\ref{fig:misclass_GY_DDC}
and~\ref{fig:MSE_GY_DDC}. 
In the top panels we see that the new method performs about as 
well as 2SGS when the correlations between the variables are
fairly low, but does much better when there are some higher
correlations.
For rowwise outliers their performance is quite similar.

\cite{Agostinelli:2SGS} showed that 2SGS is consistent,
that is, it gets the right answer for data generated from
the model without any contamination when $n$ goes to
infinity. 
The proof follows from the fact that the fraction of cells 
flagged by the univariate GY filter goes to zero in that 
setting.
This is not true for DDC because the cutoff value
\eqref{eq:qCut} is fixed, so some cells will still be flagged.
Nevertheless the above simulations indicate that
with actual contamination, using DDC 
in the first step often does better than GY.

A limitation of the GSE in the second step is that it
requires $n > d$ and in fact the dimension $d$ should not be 
above 20 or 30, whereas the raw DDC method can deal with 
higher dimensions as we saw in the glass example with 
$d=750$.

\section{\normalsize{CONCLUSIONS AND OUTLOOK}}
\label{sec:conclusions}
The proposed method detects outliers at the level of the
individual cells (entries) of the data matrix, unlike
existing outlier detection methods which consider the 
rows (cases) of that matrix as the basic units.
Its main construct is the prediction of each cell.
This turns a high dimensionality into a blessing instead
of a curse, as having more variables (columns) available 
increases the amount of information and may improve the 
accuracy of the predicted cells. 

The new method requires more computation than considering 
each variable in isolation, but on the other hand is able 
to detect outliers that would otherwise remain hidden
as we saw in the first example.
In simulations we saw that {\it DetectDeviatingCells}
performs as well as columnwise outlier detection when
there is little correlation between the columns, whereas
it does much better when there are higher correlations.
Also, using it as the first step in the method 
of~\cite{Agostinelli:2SGS} outperforms a columnwise start.

A topic for further research is to extend this work to
non-numeric variables, such as binary and nominal. 
For the interaction between numeric and non-numeric 
variables, and between non-numeric variables, this 
necessitates replacing correlation by other measures of 
bivariate association. 
Also, for predicting cells
the linear regression would need to be replaced by other 
techniques such as logistic regression. 

\section{\normalsize{SOFTWARE AVAILABILITY}}
\label{sec:software}
The R code of the {\it DetectDeviatingCells} 
algorithm is available on CRAN in the package 
{\it cellWise}, which also contains functions for drawing
cell maps and allows to reproduce all the examples in
this paper.
Equivalent MATLAB code is available from the website
{\it http://wis.kuleuven.be/stat/robust/software}\;.

\vskip0.3in

\noindent{\bf APPENDIX}

\vskip0.2in	

The building blocks of {\it DetectDeviatingCells} are some simple 
existing robust methods for univariate and bivariate data.
Several are available, and our choice was based on a 
combination of robustness and computational efficiency, as
all four of them only require $O(n)$ computing time and memory.

For estimating location and scale of a single data column
we use the first step of an algorithm for M-estimators,
as described on pages 39--41 of~\citep{Maronna:RobStat}.
In particular, for estimating location we employ Tukey's 
biweight function
$$W(t) = \left(1 - \left(\frac{t}{c}\right)^2
                    \right)^2\;I(|t| \leqslant c)$$
where $c > 0$ is a tuning constant (by default $c = 3$).
Given a univariate dataset $Y = \{y_1,\hdots,y_n\}$ we
start from the initial estimates
  $$m_1 = \med_{i=1}^{n}(y_i) \quad \mbox{  and  } \quad 
    s_1 = \med_{i=1}^{n}\,|y_i - m_1|$$
and then compute the location estimate
$$robLoc(Y) = \left(\sum_{i=1}^n w_i y_i\right)/
              \left(\sum_{i=1}^n w_i\right)$$
where the weights are given by 
$w_i = W((y_i - m_1)/s_1)$\;. 

For estimating scale we assume that $Y$ has already
been centered, e.g. by subtracting $robLoc(Y)$, so that
we only need to consider deviations from zero.
We now use the function
$\rho(t) = \mbox{min}(t^2,b^2)$
where $b = 2.5$.
Starting from the initial estimate $s_2 = \med_i(|y_i|)$
we then compute the scale estimate
   $$robScale(Y) = s_2 \sqrt{\frac{1}{\delta}\;
   \ave_{i=1}^n \rho\left(\frac{y_i}{s_2}\right)}$$
where the constant $\delta=0.845$ ensures consistency
for Gaussian data.

The next methods are bivariate, i.e. they operate on two data 
columns, call them $j$ and $h$. For correlation we start from
the initial estimate
   $$\hat{\rho}_{jh} = \left((robScale_i(z_{ij}+z_{ih}))^2 -
         (robScale_i(z_{ij}-z_{ih}))^2\right)/4 $$
\citep{GK:1972}
which is capped to lie between -1 and 1 (this assumes that
the columns of the matrix $\bZ$ were already centered at 0 
and normalized). 
This $\hat{\rho}_{jh}$ implies a tolerance ellipse around (0,0)
with the same coverage probability $p$ as in~\eqref{eq:qCut}.
Then {\it robCorr} is defined as the plain product-moment
correlation of the data points $(z_{ij},z_{ih})$ inside the 
ellipse.

For the slope we again assume the columns were already centered,
but they need not be normalized. The initial slope estimate is
 $$b_{jh} = \med_{i=1}^{n} \left(\frac{z_{ij}}{z_{ih}}\right)$$
where fractions with zero denominator are first taken out.
For every $i$ we then compute the raw residual 
   $r_{ijh} = z_{ij} - b_{jh}\,z_{ih}\;$. 
Finally we compute the plain least squares regression line
without intercept on the points for which
$|r_{ijh}| \leqslant c\, \mbox{\it robScale}_{i'}(r_{i'jh})$ where $c$ is the constant~\eqref{eq:qCut}. 
We then define {\it robSlope} as the slope of that line.
	

\vskip0.2in	

\begin{thebibliography}{}

\bibitem[{Agostinelli et~al.(2015)}]{Agostinelli:2SGS}
  Agostinelli, C., Leung, A., Yohai, V.J., 
	and Zamar, R.H. (2015),
	``Robust estimation of multivariate location and scatter
	in the presence of cellwise and casewise contamination,''
	{\it Test}, 24, 441--461.
	
\bibitem[{Alfons(2016)}]{Alfons:robustHD}
  Alfons, A. (2016)
  Package robustHD, R-package version 0.5.1,\\ 
	{\it http://CRAN.R-project.org/package=robustHD}\;.
		
\bibitem[{Alqallaf et~al.(2009)}]{Alqallaf:ICM}
  Alqallaf, F., Van Aelst, S., Yohai, V., 
	and Zamar, R.H. (2009),
  ``Propagation of outliers in multivariate data,''
  {\it The Annals of Statistics}, 37, 311--331.

\bibitem[{Arya et~al.(1999)}]{Arya:ANN}
  Arya, S., Mount, D.M., Netanyahu, N.S., Silverman, R., 
	and Wu, A.Y. (1999),
	``An optimal algorithm for approximate nearest neighbor
	searching in fixed dimensions,''
	{\it Journal of the ACM}, 45, 891--923.

\bibitem[{Bini and Bertacci(2006)}]{Bini:transformation}
  Bini, M., and Bertacci, B. (2006),
	``Robust transformation of proportions using the forward 
	search,'' in
	{\it Data Analysis, Classification and the Forward Search},
	eds. S. Zani, A. Cerioli, M. Riani, and M. Vichi,
	Springer, pp 173--180.

\bibitem[{Cheng and Victoria-Feser(2002)}]{Cheng:missing}
  Cheng, T.C. and Victoria-Feser, M.P. (2002),
	``High-breakdown estimation of multivariate mean and
	covariance with missing observations,''
	{\it British Journal of Mathematical and Statistical 
	Psychology}, 55, 317--335.

\bibitem[{Danilov(2010)}]{Danilov:PhD}
  Danilov, M. (2010),
  ``Robust estimation of multivariate scatter in non-affine 
	equivariant scenarios,'' Ph.D. dissertation, University of 
	British Columbia, Vancouver.

\bibitem[{Danilov et~al.(2012)}]{Danilov:GSE}
  Danilov, M., Yohai, V.J., and Zamar, R.H. (2012),
  ``Robust estimation of multivariate location and scatter
	in the presence of missing data,''
  {\it Journal of the American Statistical Association}, 
	107, 1178--1186.

\bibitem[{Dempster et~al.(1977)}]{Dempster:EM}
  Dempster, A., Laird, N., and Rubin, D. (1977),
	``Maximum likelihood from incomplete data via the {EM}
	algorithm,''
  {\it Journal of the Royal Statistical Society Series B},
	39, 1--38.
	
\bibitem[{Gervini and Yohai(2002)}]{Gervini:filter}
  Gervini, D., and Yohai, V.J. (2002), 
  ``A class of robust and fully efficient regression 
	estimators,''
	{\it The Annals of Statistics}, 30, 583--616.

\bibitem[{Gnanadesikan and Kettenring(1972)}]{GK:1972}
  Gnanadesikan, R., and Kettenring, J.R. (1972),
	``Robust estimates, residuals, and outlier detection with
	multiresponse data,'' {\it Biometrics}, 28, 81--124.

\bibitem[{Hubert et~al.(2005)}]{Hubert:ROBPCA}
  Hubert, M., Rousseeuw, P.J., and Vanden Branden, K. (2005),
	``ROBPCA: A new approach to robust principal component
	analysis,''	{\it Technometrics}, 47, 64--79.

\bibitem[{Human Mortality Database(2015)}]{HMD:HMD}
Human Mortality Database. University of California, Berkeley
(USA), and Max Planck Institute for Demographic Research 
(Germany). Available at {\it www.mortality.org} (data 
downloaded in November 2015).

\bibitem[{Hyndman and Shang(2010)}]{Hyndman:boxplot}
  Hyndman, R.J., and Shang, H.L. (2010),
	``Rainbow plots, bagplots, and boxplots for functional data,''
  {\it Journal of Computational and Graphical Statistics},
	19, 29--45.

\bibitem[{Kriegel et~al.(2009)}]{Kriegel:SOD}
  Kriegel, H.-P., Kr\"oger, P., Schubert, E., 
	and Zimek, A. (2009),
  ``Outlier detection in axis-parallel subspaces of high
	dimensional data,''
	in {\it Lecture Notes in Computer Science Vol. 5476},
	eds. T. Theeramunkong, B. Kijsirikul, N. Cersone,
	and T.-B. Ho,	Springer, pp 831--838.

\bibitem[{Lemberge et~al.(2000)}]{Lemberge:glass}
  Lemberge, P., De Raedt, I., Janssens, K.H., Wei, F., 
	and Van Espen, P.J. (2000),
  ``Quantitative Z-analysis of 16th-17th century
	archaeological glass vessels using {PLS} regression
  of {EPXMA} and $\mu-${XRF} data,''
	{\it Journal of Chemometrics}, 14, 751--763.

\bibitem[{Leung et~al.(2016)}]{Leung:regression}
  Leung, A., Zhang, H., and Zamar, R. (2016),
	``Robust regression estimation and inference in the
	presence of cellwise and casewise contamination,''
	{\it Computational Statistics and Data Analysis},
	99, 1--11.

\bibitem[{Little(1988)}]{Little:robust}
  Little, R.J.A. (1988),
	``Robust estimation of the mean and covariance matrix
	from data with missing values,''
  {\it Journal of the Royal Statistical Society Series C},
	37, 23--38.

\bibitem[{Lopuha\"a and Rousseeuw(1991)}]{Lopuhaa:affine}
  Lopuha\"a, H.P., and Rousseeuw, P.J. (1991),
	``Breakdown points of affine equivariant estimators of
	multivariate location and covariance matrices,''
	{\it The Annals of Statistics}, 19, 229--248.

\bibitem[{Maronna et~al.(2006)}]{Maronna:RobStat}
  Maronna, R.A., Martin, R.D., and Yohai, V.J. (2006), 
  {\it Robust Statistics: Theory and Methods},
  New York: Wiley.

\bibitem[{\"Ollerer et~al.(2016)}]{Oellerer:ShootingS}
  \"Ollerer, V., Alfons, A., and Croux, C. (2016),
  ``The shooting S-estimator for robust regression,''
  {\it Computational Statistics}, to appear.

\bibitem[{Riani(2008)}]{Riani:transformation}
  Riani, M. (2008),
	``Robust transformations in univariate and multivariate
	time series,''
	{\it Econometric Reviews}, 28, 262--278.
	
\bibitem[{Rousseeuw(1985)}]{Rousseeuw:HBD}
  Rousseeuw, P.J. (1985),
	``Multivariate estimation with high breakdown point,''
	in {\it Mathematical Statistics and Applications, Vol. B},
	eds. W. Grossmann, G. Pflug, I. Vincze, and W. Wertz,
	Dordrecht: Reidel, pp. 283--297.
	
\bibitem[{Rousseeuw and Leroy(1987)}]{Rousseeuw:RobReg}
  Rousseeuw, P.J., and Leroy, A.M. (1987), 
  {\it Robust Regression and Outlier Detection},
  New York: Wiley.

\bibitem[{Rousseeuw and Van Driessen(1999)}]{Rousseeuw:FastMCD}
  Rousseeuw, P.J., and Van Driessen, K. (1999),
	``A fast algorithm for the Minimum Covariance Determinant
  estimator,''
	{\it Technometrics}, 41, 212--223.

\bibitem[{St\"adler et~al.(2014)}]{Staedler:PA}
  St\"adler, N., Stekhoven, D.J., and B\"uhlmann, P. (2014),
	``Pattern alternating maximization algorithm for missing
	data in high-dimensional problems,''
	{\it Journal of Machine Learning Research}, 15, 
	1903--1928.

\bibitem[{Stahel and M\"achler(2009)}]{Stahel:wheel}
  Stahel, W.A., and M\"achler, M. (2009), 
	``Comment on invariant co-ordinate selection,'' 
	{\it Journal of the Royal Statistical Society Series B}, 
	71, 584--586.

\bibitem[{Tibshirani(1996)}]{Tibshirani:Lasso}
  Tibshirani, R. (1996),
  ``Regression shrinkage and selection via the {L}asso,''
  {\it Journal of the Royal Statistical Society Series B},
  58, 267--288.

\bibitem[{Van Aelst et~al.(2012)}]{VanAelst:huberized}
  Van Aelst, S., Vandervieren, E., and Willems, G. (2012), 
  ``A {S}tahel-{D}onoho estimator based on huberized 
	outlyingness,''
  {\it Computational Statistics and Data Analysis}, 56,
	531--542.

\bibitem[{Vanderkam et~al.(2013)}]{Vanderkam:Google}
  Vanderkam, S., Schonberger, R., Rowley, H., 
	and Kumar, S. (2013)
  ``Nearest Neighbor Search in Google Correlate,''
  Google Technical Report 41694,
  {\it http://www.google.com/trends/correlate/nnsearch.pdf}\;.
	
\bibitem[{Yeo and Johnson(2000)}]{Yeo:transform}
  Yeo, I.K., and Johnson, R.A. (2000), 
  ``A new family of power transformations to improve 
	normality or symmetry,'' {\it Biometrika}, 87, 954--959.

\end{thebibliography}
\end{document}